\documentclass[sigconf,nonacm]{acmart}

\usepackage{balance}
\usepackage{enumitem}
\usepackage{mathtools}
\usepackage{multirow}
\usepackage{graphicx}
\usepackage{subcaption}

\def\bitcoinA{%
  \leavevmode
  \vtop{\offinterlineskip 
    \setbox0=\hbox{B}%
    \setbox2=\hbox to\wd0{\hfil\hskip-.03em
    \vrule height .3ex width .15ex\hskip .08em
    \vrule height .3ex width .15ex\hfil}
    \vbox{\copy2\box0}\box2}}

\makeatletter
\newcommand\footnoteref[1]{\protected@xdef\@thefnmark{\ref{#1}}\@footnotemark}
\makeatother

\copyrightyear{2023}
\acmYear{2023}
\setcopyright{rightsretained}
\acmConference[CIKM '23]{Proceedings of the 32nd ACM International Conference on Information and Knowledge Management}{October 21--25, 2023}{Birmingham, United Kingdom} \acmBooktitle{Proceedings of the 32nd ACM International Conference on Information and Knowledge Management (CIKM '23), October 21--25, 2023, Birmingham, United Kingdom}\acmDOI{10.1145/3583780.3614790  }
\acmISBN{979-8-4007-0124-5/  23/10}

\makeatletter
\gdef\@copyrightpermission{
  \begin{minipage}{0.3\columnwidth}
   \href{https://creativecommons.org/licenses/by-nc/4.0/}{\includegraphics[width=0.90\textwidth]{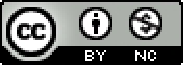}}
  \end{minipage}\hfill
  \begin{minipage}{0.7\columnwidth}
   \href{https://creativecommons.org/licenses/by-nc/4.0/}{This work is licensed under a Creative Commons Attribution-NonCommercial International 4.0 License.}
  \end{minipage}
  \vspace{5pt}
}
\makeatother

\begin{document}

\title{Beyond Trading Data: The Hidden Influence of Public Awareness and Interest on Cryptocurrency Volatility} 



\author{Zeyd Boukhers}
\orcid{0000-0001-9778-9164}
\affiliation{%
  \institution{Fraunhofer Institue for Applied Information Technology\\
  University of Koblenz}
  \streetaddress{Schloss Birlinghoven}
  \country{Germany}
  \postcode{53757}
}
\email{zeyd.boukhers@fit.fraunhofer.de}

\author{Azeddine Bouabdellah}
\orcid{0009-0003-4413-0042 }
\affiliation{%
  \institution{University of Koblenz}
  \streetaddress{Universitätsstraße 1}
  \city{Koblenz}
  \country{Germany}}
\email{bouabdallahazeddine@gmail.com}

\author{Cong Yang}
\orcid{0000-0002-8314-0935}
\affiliation{%
  \institution{Soochow University}
  \city{Suzhou}
  \country{China}
}
\email{cong.yang@suda.edu.cn}

\author{Jan Jürjens}
\orcid{0000-0002-8938-0470}
\affiliation{%
  \institution{University of Koblenz\\
  Fraunhofer Institute for Software and Systems Engineering
  }
  \streetaddress{Universitätsstraße 1}
  \country{Germany}}
\email{juerjens@uni-koblenz.de}

\renewcommand{\shortauthors}{Zeyd Boukhers, Azeddine Bouabdellah, Cong Yang, \& Jan Jürjens} 
\begin{abstract}
Since Bitcoin first appeared on the scene in 2009, cryptocurrencies have become a worldwide phenomenon as important decentralized financial assets. Their decentralized nature, however, leads to notable volatility against traditional fiat currencies, making the task of accurately forecasting the crypto-fiat exchange rate complex. This study examines the various independent factors that affect the volatility of the Bitcoin-Dollar exchange rate. To this end, we propose \emph{CoMForE}, a multimodal AdaBoost-LSTM ensemble model, which not only utilizes historical trading data but also incorporates public sentiments from related tweets, public interest demonstrated by search volumes, and blockchain hash-rate data. Our developed model goes a step further by predicting fluctuations in the overall cryptocurrency value distribution, thus increasing its value for investment decision-making. We have subjected this method to extensive testing via comprehensive experiments, thereby validating the importance of multimodal combination over exclusive reliance on trading data. Further experiments show that our method significantly surpasses existing forecasting tools and methodologies, demonstrating a 19.29\% improvement. This result underscores the influence of external independent factors on cryptocurrency volatility. 

\end{abstract}






\maketitle

\section{Introduction}
\label{sec:intro}

Since the mining of the first Bitcoin's genesis block in 2009, cryptocurrencies have redefined the landscape of financial assets. By January 2022, the total market capitalization of major cryptocurrencies nearly reached an astounding \$1 trillion\footnote{https://gadgets.ndtv.com/cryptocurrency/news/bitcoin-price-btc-cryptocurrency-market-crash-usd-1-trillion-coinmarketcap-2726233}. The global cryptocurrency market exhibits a Compound Annual Growth Rate (CAGR) of 30\% from 2019 to 2026\footnote{https://www.globenewswire.com/news-release/2021/04/12/2208331/0/en/At-30-CAGR-CryptoCurrency-Market-Cap-Size-Value-Surges-to-Record-5-190-62-Million-by-2026-Says-Facts-Factors.html}. This remarkable growth not only offers lucrative investment and trading opportunities but also contributes to further expansion of the cryptocurrency market, fostering a virtuous cycle. In contrast to traditional FIAT currencies, which are regulated by central banks and governing bodies, cryptocurrencies are entirely decentralized. Transactions are validated and processed via a cryptographic network of nodes and are logged on a blockchain -- a digital transaction ledger~\cite{chuen2017cryptocurrency}. These unique properties make forecasting the fiat-crypto exchange rate (or simply, cryptocurrency price forecasting) an exceptionally complex and often error-prone task~\cite{abraham2018}. Given this complexity, financial analysts and AI experts continually dissect the market to deepen their understanding of price fluctuation trends~\cite{aboody2018, abraham2018, alkhodhairi2021, kristoufek2015, chevallier2021, jang2017, kumar2020}.

Recent studies focused on forecasting cryptocurrency prices have predominantly leveraged neural networks due to their exceptional performance in related tasks~\cite{kumar2020, livieris2020, nasekin2019,pintelas2020,yiying2019}. As a result, these approaches have achieved superior outcomes compared to traditional machine learning and statistical methods~\cite{yang2019, kumar2020}. However, a significant limitation of existing methods is that they consider only a handful of factors influencing the cryptocurrency market. The common practice is to infer a predictive function from available training sets and then evaluate the derived functions based on their generalization capabilities, presuming that price series often display homogeneous nonstationarity~\cite{pintelas2020}.

In reality, cryptocurrency volatility is influenced by a multitude of factors, distinct from those affecting foreign exchange rates. Factors such as the cryptocurrency mining process's hash rate and public awareness play a critical role in price volatility~\cite{kristoufek2015}. For example, studies have revealed a sensitivity of cryptocurrency prices to public opinion sentiments~\cite{aboody2018,lamon2017}. Yang et al.~\cite{yang2019} proposed that social media sentiment serves as a valuable predictor of future Bitcoin price volatility. Akbiyit et al.~\cite{akbiyik2023ask} have proved this hypothesis in their study. Therefore, a holistic approach that considers these wider influences is essential for understanding the volatility of cryptocurrencies.

To enhance forecasting accuracy, this paper proposes a comprehensive utilization of all available factors that influence cryptocurrency prices. Specifically, we introduce \emph{CoMForE}: \textit{\textbf{Co}mprehensivle \textbf{M}ultimodal Crypto \textbf{For}ecasting \textbf{E}nsembler}, an ensemble of multimodal Long Short-Term Memory (LSTM) models for cryptocurrency volatility prediction. This method employs trading data, social media sentiment analysis, blockchain data (including hash rate and network difficulty), and search volumes from search engines. Our focus is on Bitcoin due to its dominant market presence, its popularity among the 7812 existing cryptocurrencies, and the ample data available for analysis. Ultimately, our goal is to facilitate sound investment decision-making through the provision of reliable price forecasts and volatility distribution assessments. The following research questions guide our exploration:

\begin{itemize}[leftmargin=*]
    \item \textbf{RQ1:} Which combinations of data modalities exert the most significant influence on cryptocurrency volatility?
    \item \textbf{RQ2:} Does the application of ensemble learning to multimodal data yield effective outcomes in the forecasting of cryptocurrency prices?
    \item \textbf{RQ3:} How can we effectively interpret cryptocurrency price forecasting in conjunction with volatility distribution?
\end{itemize}

\noindent To this end, the main contributions of this paper can be summarized as follows:

\begin{itemize}[leftmargin=*]
\item Our research encompasses factors influencing cryptocurrency volatility, including trading data, social media sentiments, blockchain metrics, and search engine volumes.
\item We introduce an LSTM-based multimodal ensemble learning model that outperforms existing models, delivering reliable decision-making support to investors.
\item Beyond price prediction, our model provides fluctuation distribution, enhancing investors' understanding of forecast certainty.
\item Comprehensive experiments and analyses affirm the robustness of our approach.
\item We provide an open-source implementation of our model for further development and improvement.
\end{itemize}


\section{Related Work}
\label{sec:rel-work}

In this section, we review the related works divided into three categories: 

\subsection{Traditional Market Price Forecasting}

The task of forecasting stock prices has long been a subject of interest for scholars in the realms of finance, statistics~\cite{hansen1999time}, and data science~\cite{yoo2021accurate,li2019multi}. The aim is to empower investors with the tools necessary to augment their profits while mitigating potential losses. One of the early approaches adopted was the application of statistical models. An in-depth comparative study was conducted by Haviluddin et al.~\cite{alfred2015performance}, examining statistical and machine learning techniques for short-term forecasting using time-series data. The study primarily compared the statistical method ARIMA, Neural Networks, and genetic algorithms. The findings underscored the superior efficiency and reliability of Neural Networks for short-term time series forecasting.

Over the past decade, the advancements in deep neural networks across diverse applications have brought them into the spotlight for financial and economic forecasting, including time-series predictions~\cite{hossain2018hybrid, selvin2017stock, khare2017short, nikou2019stock}. Deep learning approaches have demonstrated their ability to significantly outperform other methods, largely due to their capability to learn concealed features of time series and historical market trends. Traditional markets are governed by relatively well-understood factors and rules that directly influence price trends, such as the number of asks, bids, and transactions. This relative clarity facilitates deep learning models in pattern recognition from historical data, leading to highly reliable forecasting, as evidenced by the high accuracy and low error rates observed when applying price, ask, and bid time series alone~\cite{nikou2019stock, selvin2017stock}.

However, attempts to apply these cutting-edge deep learning methods—proven effective in traditional markets—to the task of forecasting cryptocurrency prices have not yielded satisfactory results~\cite{madan2015automated}. The distinct characteristics of the cryptocurrency market render it a more complex domain for forecasting.

\subsection{Machine Learning for Cryptocurrency Price Forecasting}

Cryptocurrency market volatility has driven the development of tailored forecasting approaches. Yiying and Yaze~\cite{yiying2019} proposed LSTM and ANN architectures using price, ask, and bid time series for short and long-term price predictions of Bitcoin, Ethereum, and Ripple. The results showed the LSTM's effectiveness in capturing short-term dynamics.

McNally et al.~\cite{mcnally2018predicting} evaluated LSTM and Bayesian-optimized Recurrent Neural Network models for Bitcoin price forecasting, concluding that the LSTM had marginally superior accuracy. However, they noted the challenge of balancing overfitting and underfitting due to the high volatility of cryptocurrency time series.

Kumar and Rath~\cite{kumar2020} used only historical trading data for Ethereum price predictions, indicating LSTM's slight outperformance over MLP for short-term forecasts. Meanwhile, Pintelas et al.~\cite{pintelas2020} found LSTM-based and CNN-based models demonstrating almost random walk processes, suggesting the exploration of new approaches.

Chevallier et al.~\cite{chevallier2021} introduced an AdaBoost-based approach for cryptocurrency forecasting that significantly improved performance. Despite its simplicity, AdaBoost outperformed other models like ANNs, LSTMs, KNNs, and SVMs, highlighting its generalizability and interpretability.

\subsection{Sentiment Analysis and Multimodality for Cryptocurrency Price Forecasting}

Krisoufek~\cite{kristoufek2015} found that the cryptocurrency market is influenced by various factors, including the number of asks, bids, exchange rates, and notably, public awareness, as confirmed by Akbiyik et al.\cite{akbiyik2023ask}. Recognizing the challenge of quantifying public awareness, individual sentiments can serve as proxies, which are believed to affect price trends\cite{schinckus2020good}. Young et al.~\cite{kim2016predicting} proposed that cryptocurrency forum sentiments might influence Bitcoin prices. They proposed a model using exclusively sentiment data and found a correlation between price trends and forum sentiments. 
Leveraging public opinion for finance is not a new concept, and it extends beyond cryptocurrency. Several approaches have been suggested to utilize public sentiment for predicting stock market trends~\cite{bollen2011twitter, nguyen2015topic, xu2018stock}. However, for cryptocurrencies, it tends to have an influence rather than a correlation. 

Following these insights, several studies~\cite{nasekin2019, abraham2018, huang2021lstm} integrated sentiment analysis with LSTM models to predict next-day trading prices. They gathered posts related to cryptocurrencies, assigned each post a sentiment score, and merged these scores with trading data into a singular vector for the LSTM model. Their results revealed a modest enhancement in forecasting outcomes compared to models relying solely on trading data. Furthermore, Huang et al.~\cite{huang2021lstm} supported their approach with an autoregressive model that improved accuracy and recall by 18.5\%.

However, these studies~\cite{nasekin2019, abraham2018, huang2021lstm, chevallier2021, pintelas2020, kumar2020} overlooked the influence of interaction levels on posts' impact such as likes, comments, shares, and exclusively focused on online communities as fluctuation factors, neglecting other crucial factors~\cite{kristoufek2015}. 

In addition to the findings of Krisoufek~\cite{kristoufek2015}, the other modalities have proven to be correlated with the stock market, such as search volume~\cite{da2011search,swamy2019investor}
\section{Approach: \emph{CoMForE}}
\label{sec:app}
\begin{figure}
    \centering
    \includegraphics[width= 0.9\linewidth]{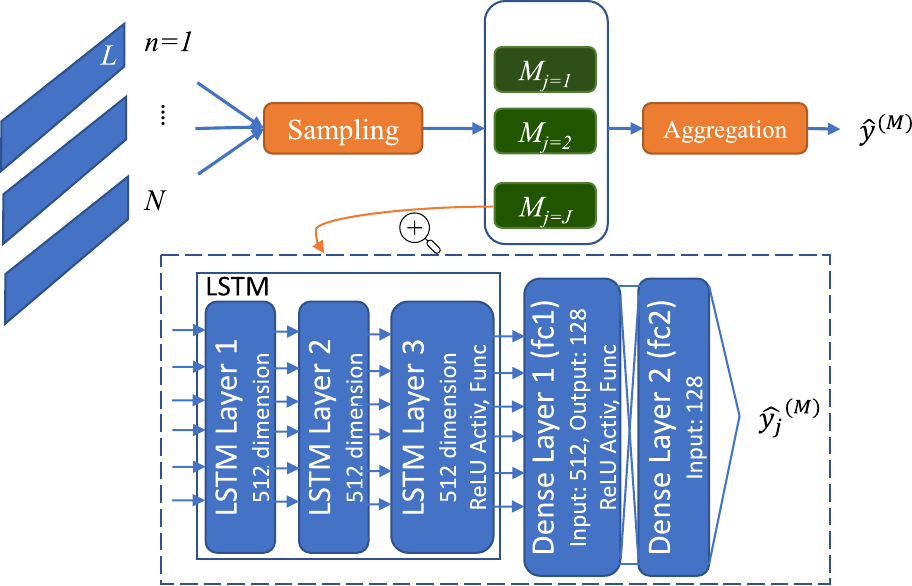}
    \caption{Overview of \emph{CoMForE}'s Architecture for Cryptocurrency Price Prediction}
    \label{fig:app:ovr}
\end{figure}

To address our first research question (\textbf{RQ1}), this paper proposes a novel strategy for forecasting next-day cryptocurrency volatility by leveraging all indicative factors. To tackle \textbf{RQ2}, we introduce an ensemble learning approach that marries adaptive boosting with LSTM architectures, known for their proficiency in learning hidden trends within sequential data, especially for short-term forecasting. As depicted in Figure~\ref{fig:app:ovr}, our approach comprises multiple LSTM weak learners (i.e., forecasters) that collaboratively construct a cryptocurrency volatility prediction model. Each LSTM model $j$ is trained on a subset sampled from the original dataset and subsequently assigned a weighted score $w_j$ based on its performance during the inference phase. Each subset is sampled using the sampling weights $s_{n=1}^{N}$ for the $N$ samples. Here, a sample denotes the input sequence ${\textbf{x}}_{t-k}^{t}$ associated with its respective price, where $k$ is the length of the input layer and $t$ is the time stamp (i.e. day). Initially, the sampling weights denoted as $s_{n=1}^{N}$ are set to $\frac{1}{N}$, indicating that all samples carry equal importance in the learning process. Then, for each LSTM forecaster $M_j$, the following steps are performed:

\begin{enumerate}[leftmargin=*]
\item Randomly select $1<l<=L$ features from the total $L$ features
\item Train the LSTM model $M_j$ using the sampled subset. This involves learning the weights $W_{f,j}$, $W_{i,j}$, $W_{C,j}$, $W_{o,j}$ and biases $b_{f,j}$, $b_{i,j}$, $b_{C,j}$, $b_{o,j}$ of the LSTM.
\item Calculate the errors $e_{n=1:N}^{(j)}$ of the LSTM forecaster $M_j$ for each sample $n$.
\item Compute the total error $E_j$ of the $j$-th LSTM forecaster, such that:
\begin{equation}
E_j = \sum_{n=1}^{N} s_{n}^{(j)} e_n^{(j)}
\end{equation}
\item Calculate the weight $w_j$ of the $j$-th forecaster as follows:
\begin{equation}
w_j = \log \left(\frac{1 - E_j}{E_j}\right).
\end{equation}
\item Update the sampling weights $s_{n}^{(j)}$ of all samples as follows:
\begin{equation}
s_{n}^{(j)} = \frac{s_{n}^{(j)} \exp \left( w_j I\left( y_n \neq M_j(x_n) \right) \right)}{Z_j},
\end{equation}
where $Z_j$ is a normalization factor making $s_{n}^{(j)}$ a distribution.
\end{enumerate}

The normalization factor $Z_j$ ensures that the updated weights $s_{n}^{(j)}$ form a valid probability distribution. The indicator function $I\left( y_n \neq M_j(x_n) \right)$ checks if the predicted output of the $j$-th model for the $n$-th sample does not match the true output for this sample. If true, $I$ returns 1; otherwise, it returns 0.


After all LSTM forecasters are trained, we calculate the final ensemble prediction $\hat{y}_{n}^{(M)}$ for a given sample $n$ as follows:


\begin{equation}
\hat{y}_{n}^{(M)} = \frac{\sum_{j=1}^{J} \hat{y}_{n}^{(j)} w_j}{\sum_{j=1}^{J} w_j},
\end{equation}

where $J$ signifies the total number of LSTM forecasters, $\hat{y}_{n}^{(j)}$ represents the prediction made by the $j$-th LSTM model $M_j$ for the $n$-th sample, and $w_j$ denotes the weight of the $j$-th LSTM forecaster. This weight reflects the relative contribution of each forecaster's prediction towards the final ensemble prediction.

Each LSTM forecaster in our ensemble provides a one-day-ahead forecast of the price of Bitcoin (\bitcoinA) in US dollars (\$). To do this, the forecaster uses a seven-day window of historical data, denoted as ${\textbf{x}}_{t-7}^{t}$. Here, $\textbf{x}_t$ represents a normalized feature vector of length 18 corresponding to day $t$. This vector combines representations from four distinct modalities: trading data, sentiments, blockchain data, and search volumes. Each of these modalities contributes to the richness and complexity of the feature space and is described in greater detail as follows:

\subsection{Input modalities}

While there is a multitude of cryptocurrencies available, Bitcoin (\bitcoinA) remains the most prominent, and data related to it is readily accessible. Therefore, this study centres on Bitcoin as the primary use case. We have gathered a comprehensive dataset spanning from 2012 through the end of 2020, comprising the following modalities:

\subsubsection{Trading data:}

Trading data encapsulates a sequential collection of attributes that depict the Bitcoin market's dynamics. Each timestamp $t$ corresponds to a single day, with the opening price recorded at 00:00 and the closing price at 23:59. At each $t$, the following market characteristics are noted:

\begin{itemize}[leftmargin=*]
\item \emph{Open Price (\$)}: The price of one Bitcoin (\bitcoinA) at the beginning of $t$.
\item \emph{High Price (\$)}: The peak price of one \bitcoinA~ reached within the period of $t$.
\item \emph{Low Price (\$)}: The lowest price of one \bitcoinA~ recorded within the period of $t$.
\item \emph{Close Price (\$)}: The price of one \bitcoinA~ at the end of $t$.
\item \emph{Bitcoin Volume (\bitcoinA)}: The total quantity of \bitcoinA~ traded during the interval of $t$.
\item \emph{Dollar Volume (\$)}: The total value in dollars of \bitcoinA~ traded during $t$.
\item \emph{Weighted Price (\$)}: The weighted average price of \bitcoinA~ traded during $t$.
\item \emph{Average Transaction Fee (\$)}: The average transaction fees charged by the top 20 trading and exchange platforms during $t$.
\item \emph{Number of Transactions}: The cumulative number of transactions carried out during $t$.
\end{itemize}


\subsubsection{Public awarness:}

In today's world, tweets serve as an instantaneous source of news. Hence, in this study, they are perceived as a crucial indicator of cryptocurrency price trends. Accordingly, we gathered tweets responding to the wildcard queries: ``\emph{*Bitcoin*}'' and ``\emph{*BTC*}'', resulting in a corpus of over 120 million tweets. To filter out spam, tweets containing the hashtag ``\emph{\#Bitcoin}'' but not within the main text body were excluded.

The primary goal behind this data collection is to conduct sentiment analysis and incorporate the resulting sentiment scores into our model, thereby encoding the factor of public awareness. To this end, we utilized two widely recognized and publicly available sentiment analysis methodologies: \emph{Vader}~\cite{hutto2014vader} and \emph{Deeply Moving}\cite{socher2013recursive}. Both techniques yield a sentiment score $\theta_{\gamma}$ for tweet $\gamma$, ranging from $-1$ (extremely negative) to $1$ (extremely positive), while $0$ stands for neutral. The dual-method approach was chosen to balance their inherent sensitivity—intensifying the sentiment score when both methods agree and diminishing it when they disagree.

It is important to note that not all cryptocurrency-related tweets have an equal impact on the market. The key determinant here is the level of engagement (i.e., the tweet’s reach and number of interactions). According to recent statistics on Twitter engagement\footnote{\url{https://mention.com/en/reports/twitter/engagement/\#2}}, the median number of likes and comments per tweet is zero, implying a substantial portion of tweets receive little to no engagement. Thus, assigning equal sentiment weights to all tweets would not accurately represent the real-world scenario.

To address this, we propose weighting each tweet in a manner that gives higher consideration to those with substantial engagement. Owing to the complexity in determining the relative importance of different engagement factors, we opted for a straightforward approach to compute the weight $\omega_{\gamma}$ of each tweet $\gamma$ as the harmonic mean of the number of likes, comments, retweets, and quotes. Subsequently, these weights were normalized using min-max normalization. The final sentiment score $\theta_{\gamma}^*$ is then multiplied by $\omega_{\gamma}$; $\theta_{\gamma}^* = \theta_{\gamma} \times \omega_{\gamma}$.

To create a daily sequence of sentiment scores, all weighted sentiments of tweets posted on a given day $t$ are averaged, generating a singular value representing the overall sentiment on Twitter for that day.



\subsubsection{Blockchain details:}
\label{sec:3:block}
As asserted in the financial study by Kristoufek \cite{kristoufek2015}, blockchain metrics significantly influence the price dynamics of cryptocurrencies, notably \bitcoinA. As a result, our strategy incorporates several blockchain metrics that have been demonstrated to impact the market in financial research \cite{kristoufek2015}. Relying on various blockchain data sources\footnote{\label{foot:block}\href{https://www.blockchain.com/}{Blockchain.com}, \href{https://ycharts.com/}{YCharts}, \href{https://bitinfocharts.com/}{BitInfoChart}, and \href{https://data.nasdaq.com/}{Nasdaq Data Link}}, we include the following indicators for each day $t$:

\begin{itemize}[leftmargin=*]
\item \textit{Hash Rate:} An estimation of the computational speed at which the \bitcoinA~ network is operating.
\item \textit{Block Size:} The magnitude of a completed \bitcoinA~ block.
\item \textit{Block Time:} The time consumed to mine and generate a new \bitcoinA~ block.
\item \textit{Network Difficulty:} The complexity associated with mining \bitcoinA~ blocks across the network.
\item \textit{Active Addresses:} The aggregate number of functioning addresses.
\item \textit{Mining Profitability:} The projected average profit yielded from mining a single \bitcoinA~ block.
\end{itemize}

All the above-listed blockchain attributes follow a sequential pattern with a one-day gap between each recorded data point.

\subsubsection{Search volumes}
\label{sec:3:search}
Public curiosity can often be gauged through a variety of behaviours, and in today's digital age, online searches serve as a significant reflection of societal interest. Thus, we incorporate search volumes into our analysis, specifically focusing on queries involving either the term ``\emph{Bitcoin}'' or the acronym ``\emph{BTC}''. Given that Google, being utilized by 52\% of the global population\footnote{\url{https://review42.com/resources/google-statistics-and-facts/}}, is one of the most widely used search engines, we have leveraged the Google API to collect search data for the specified terms. This API returns the volume of searches conducted within a particular time span. These search volumes are then aggregated for each day and normalized, providing an indication of public curiosity on a daily basis.

\subsection{Fluctuation Analysis}
\label{subsec:fa}

Our primary objective is to augment the decision-making process for cryptocurrency investors. While price prediction is a crucial aspect, it's insufficient given the varying volatility in the cryptocurrency market. During low market fluctuation, model predictions align closely with the actual value. In contrast, during high volatility, the predicted price can significantly diverge. Existing methodologies do not encapsulate this level of fluctuation. Hence, in addressing \textbf{RQ3}, we aim to analyze these fluctuations using different variations of the model, each trained using a unique combination of modalities and dropout rates.

\subsubsection{Input varieties:}

Given the model described above, and incorporating data from \textit{Trading}, \textit{Twitter Sentiments}, \textit{Blockchain Details}, and \textit{Online Search Volumes}, we train multiple model variations. Each model variation, utilizing different combinations of modalities, generates a slightly distinct price forecast. In terms of architecture, these model variations only differ in the length of their input layer.

\begin{itemize}[leftmargin=*]

\item \textbf{Trading Data:} The fundamental input for cryptocurrency price forecasting, consisting of eight essential features: \textit{["Open", "High", "Low", "Close", "Volume BTC", "Volume Currency", "Weighted Price", "Average Fees"].}

\item \textbf{Twitter Sentiments:} This variant of our model uses only Twitter sentiments, captured through "Weighted Twitter Sentiments" and "Tweet Volumes".

\item \textbf{Trading Data and Blockchain Details:} This model variant incorporates trading data combined with blockchain details: \textit{["Hash Rate", "Block Size", "Block Time", "Network Difficulty", "Number of Active Addresses", "Mining Profitability"]}.

\item \textbf{Trading Data and Search Volumes:} This model variant merges trading data with \textit{["Online Search Volumes (Google Searches)"]} to improve its predictive power.

\end{itemize}

\subsubsection{Model dropouts:}

To create variability in our model's predictions and thus estimate an output distribution, we introduce different dropout rates at various layers of the trained model. This method also serves as a regularization technique to counter overfitting. Here are the variants generated with different dropout rates:

\begin{itemize}[leftmargin=*]
    \item \emph{Models \(V_1\), \(V_2\), \(V_3\)}: These models implement dropout rates of $0.1$, $0.2$, and $0.35$ respectively at the last hidden layer of the LSTM.
    \item \emph{Models \(V_4\), \(V_5\), \(V_6\)}: These variants use dropout rates of $0.1$, $0.2$, and $0.35$ respectively at the output of the fully connected layer (fc1).
\end{itemize}

\subsubsection{Predicted Price Distribution:}
We propose that the distribution of fluctuations can provide valuable insights for decision-makers regarding the most suitable actions to take. Paired with the price forecast, they can offer a measure of the forecast's uncertainty. To estimate the parameters (i.e., Mean and Variance) of the distribution based on the sample of outputs, we utilized Maximum Likelihood Estimation (MLE).

Given the outputs $O=\{o_1, o_2, …, o_{10}\}$ from the ten model variants, the aim is to estimate the parameter set $\hat{\theta}=\{\mu, \sigma^2\}$ that maximizes the likelihood function $L(\theta;O)$. This can be represented as:

\begin{equation}
L(\theta;O) = \prod_{i=1}^{10} \mathcal{N}(o_i;\theta)
\end{equation}

Then, the set of parameters that maximizes this likelihood function is given by: $\hat{\theta}=\arg\max L(\theta;O)$.

Here, $\mathcal{N}(o_i;\theta)$ denotes the normal distribution of the output $o_i$ with the parameter set $\theta$, and $\hat{\theta}$ represents the estimated parameters that maximize the likelihood function.

\section{Experiments}
\label{sec:exp}

In this section, we assess \emph{CoMForE} by comparing its performance to several other baseline approaches. To ensure reproducibility and facilitate further exploration, we have made our implementation available on GitHub\footnote{\url{https://doi.org/10.5281/zenodo.8265158}}.

\subsection{Experimental Setup}
\label{apn:exp}

Across all experiments, we maintain the same setup and parameters. These are as follows:

\begin{itemize}[leftmargin=*]
\item Loss function during training: Mean Squared Error (\emph{MSE})
\item Optimizer: \emph{Adam}, Initial learning rate: \emph{0.0003} and Number of epochs: \emph{200}
\item The training process takes place on a GPU server with the following specifications: an AMD Ryzen Threadripper 1950X 16-Core Processor, 128GiB of system memory, and an NVIDIA GV100 GPU.
\item All experiments employ the Mean Absolute Error (MAE) as the validation loss, calculated every ten epochs.
\end{itemize}

\subsection{Datasets}
\begin{figure}[h]
\includegraphics[width=0.9\linewidth]{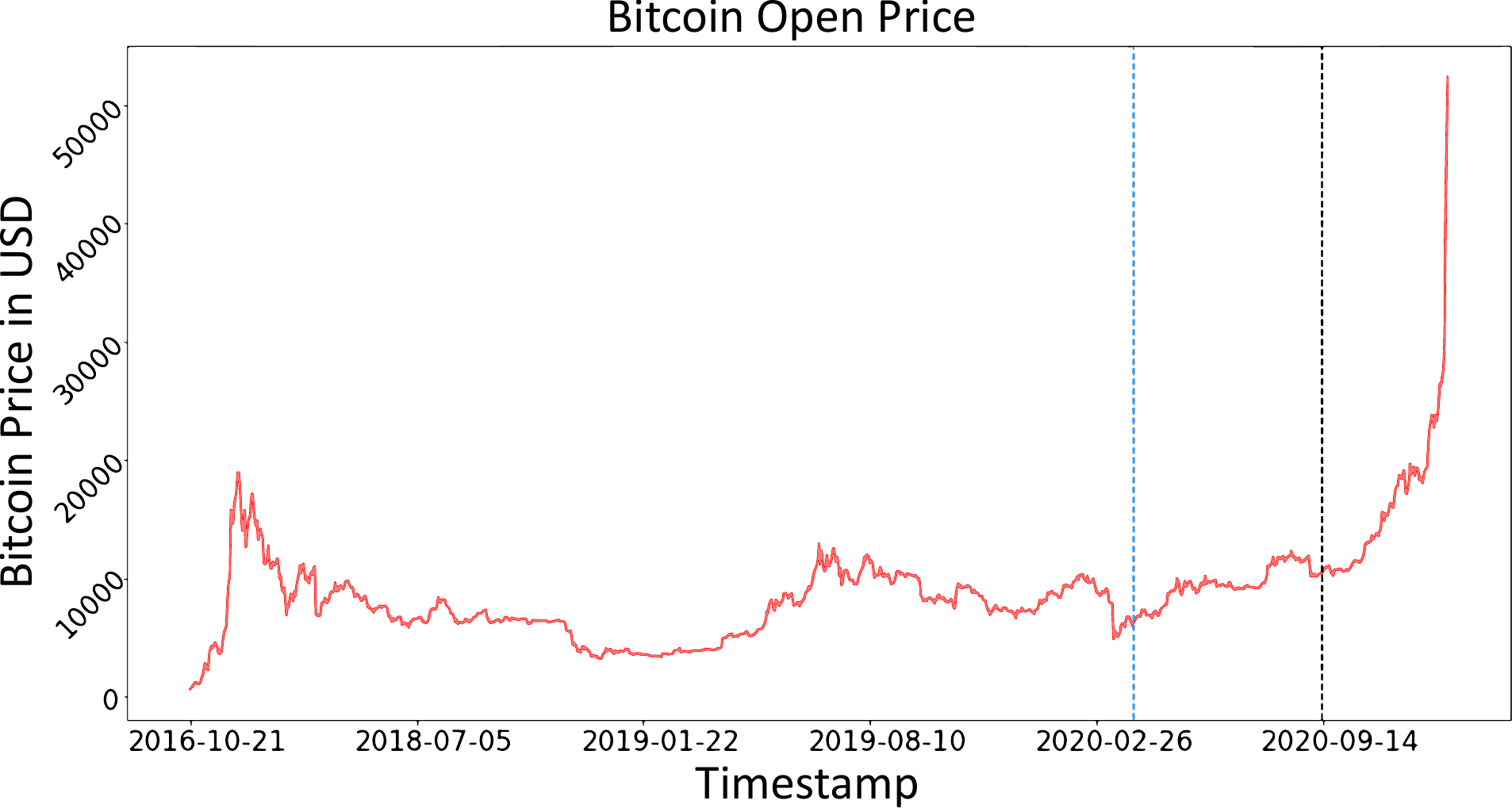}
\caption{Data split for training, validation, and testing (70\% training, 15\% validation, 15\% testing)}
\label{fig:datasplit}
\centering
\end{figure}

This study focuses on Bitcoin (\bitcoinA), the dominant cryptocurrency with readily accessible data. We gathered Bitcoin data from 2016 to 2020, yielding 1825 data points (days). The data was split into 70\% for training, 15\% for validation, and 15\% for testing, as shown in Figure \ref{fig:datasplit}. Next, we proceed to a detailed exploration of each data modality.

\subsubsection{Trading Data:}
\label{apn:data}
Acquiring comprehensive and accurate historical trading data for cryptocurrencies can be a challenging task due to the presence of missing values and occasional skipped days in many data sources. To address this, we have amalgamated data from several sources, each supplementing the gaps in the other. We also account for the potential variances in values across exchanges, given the lack of a standardized pricing protocol for cryptocurrencies. The sources for our data are as follows:

\begin{itemize}[leftmargin=*]
\item \textbf{Kaggle:} This open-source dataset comprises 1-minute interval data of \bitcoinA~ prices, collected from January 1, 2012, to March 1, 2021.

\item \textbf{Binance:} As a major cryptocurrency exchange, Binance publicly provides its data. In the absence of an API, data was manually extracted from the website. 

\item \textbf{Coinbase:} Unlike Binance, Coinbase offers APIs to gather trading data. As a major cryptocurrency exchange, its data is widely used, including by organizations like Google.

\end{itemize}

\subsubsection{Public Awareness:}

For sentiment analysis of the collected tweets, a distinct subdiscipline, this paper employs two renowned, pre-trained sentiment analysis tools: \emph{Vader}\cite{hutto2014vader} and \emph{Deeply Moving}\cite{socher2013recursive}.

\emph{Vader}~\cite{hutto2014vader}, a rule-based model tailored for social media sentiment analysis, has exhibited superior results in recent evaluations compared to various rule-based or machine learning approaches. It is expected to generalize better across different contexts. The model's output helps classify a tweet into one of nine sentiment classes, ranging from extremely negative to extremely positive.

\emph{Deeply Moving}~\cite{socher2013recursive} is another tool that examines text as a holistic entity, maintaining word correlations. This method is primarily designed for movie reviews sentiment analysis, meaning that the text is formally represented, which is different from the structure used in Twitter’s tweets, such as emojis, hashtags, and abbreviations. Despite the difficulty of fine-tuning this model on tweet sentiments due to the scarcity of labelled data, we aim to utilize both models to harness their unique strengths - Vader's explicit training on tweets and Deeply Moving's ability to comprehend sentiments of an entire sentence.

Both sentiment analysis models (Vader, Deeply Moving) are applied to all previously gathered tweets. The final sentiment score is then derived as the average of the scores from both models.

\subsubsection{Blockchain data:}
As highlighted in Section~\ref{sec:3:block}, the blockchain data has been sourced from various providers\footnoteref{foot:block}.

\subsubsection{Search volumes:}

\begin{figure}
\centering
\includegraphics[width=0.9\linewidth]{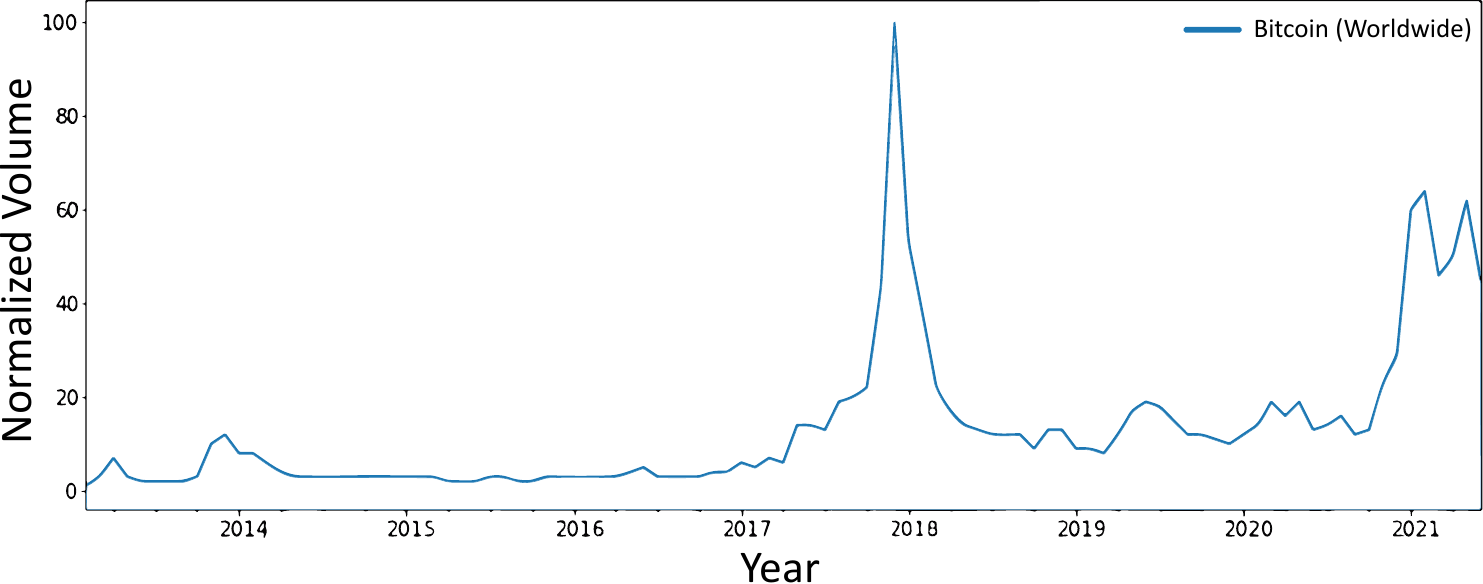}
\caption{Visualization of \textit{"Bitcoin"} search volume from 2013 to 2021.}
\label{fig:searchv}
\end{figure}

As previously stated in Section~\ref{sec:3:search}, we employed Google's search engine statistics via its API in this study. Figure \ref{fig:searchv} represents the search volumes associated with the query \textit{"Bitcoin"} from 2013 to 2021.

\subsection{Baselines}

To evaluate \emph{CoMForE}, we selected six notable and modern techniques, each with a unique architecture or different data types, to forecast Bitcoin prices. These include:

\begin{itemize}[leftmargin=*]
\item \emph{ARIMA19~\cite{alahmari2019}:} Utilizes the ARIMA model, focusing exclusively on Bitcoin trading data to forecast the next day's prices.

\item \emph{BNN17~\cite{jang2017}:} Employs Bayesian Neural Networks (BNN) to predict future Bitcoin prices, harnessing trading data.

\item \emph{LSTM20A~\cite{uras2020}:} Adopts a multivariate LSTM model tailored for cryptocurrency price predictions.

\item \emph{LSTM20B~\cite{mudassir2020}:} Uses an LSTM architecture, grounded on trading data, to predict price trajectories.

\item \emph{GRU20~\cite{dutta2020}:} Employs a Gated Recurrent Unit (GRU) infused with recurrent dropout to enhance Bitcoin price prediction accuracy.

\item \emph{GRU21~\cite{alkhodhairi2021}:} While also leveraging a Gated Recurrent Unit, this method has a different predictive focus.

\item \emph{AdaBoost21~\cite{chevallier2021}:} Leverages the traditional AdaBoost algorithm, incorporating multiple decision tree weak learners for its predictions. Its straightforward design achieves noteworthy results, even surpassing more sophisticated methods.

\end{itemize}

Since the source codes of these approaches are not publically available, we re-implemented them from scratch and reproduce their results. To ensure a fair comparison, we evaluated all the baseline methods, including \emph{CoMForE}, on the exact same dataset within the same time period.

\subsection{Results and Discussion}

\subsubsection{Modality Analysis}

To address \textbf{RQ1}, we conducted an evaluation using both \emph{CoMForE} and an \emph{LSTM}-based architecture to investigate the effectiveness of various combinations of data modalities. Both models underwent training for 200 epochs and adhered to the experimental setup outlined earlier. Note that the \emph{LSTM} model shares the same architecture as a single weak learner in \emph{CoMForE}, but trained with the full dataset. The results, displayed in Table \ref{tab:comp1}, present the performance of both models using (a) only trading data, (b) only sentiments, and (c) a combination of trading data and sentiments.  Remarkably, the results from the sentiment data alone (b) underscore the correlation between social media sentiment and cryptocurrency prices. The results derived from the combination of trading data and Twitter sentiments (c) affirm that this combination yields the most accurate results. This suggests that social media sentiments can effectively complement trading data in capturing cryptocurrency price fluctuations, thereby enhancing the model's forecasting capability. Specifically, adding sentiment analysis to both models can improve the results by up to 15\% in comparison to using only trading data.

\begin{table*}[h!]
    \centering
    \begin{tabular}{|c||c|c||c|c||c|c|}
        \hline
        \multirow{2}{*}{} & \multicolumn{2}{c||}{(a) Only Trading} & \multicolumn{2}{c||}{(b) Only Sentiments} & \multicolumn{2}{c|}{(c) Trading + Sentiments}\\
        \cline{2-7}
         & \emph{LSTM} & \emph{CoMForE}  & \emph{LSTM}   & \emph{CoMForE} & \emph{LSTM} & \emph{CoMForE} \\\hline
        Training RMSE (\$) 						& 346.507 & \textbf{100.593}* & 2,617.935 & \textbf{221.707} & 344.082 & \textbf{104.497} \\ \hline 
        Training MAE (\$) 						& 204.773 & \textbf{64.651} & 2,120.712 & \textbf{59.85}* & 209.094 & \textbf{61.134} \\ \hline
        Validation RMSE (\$) 					& \textbf{502.473} & 1,097.734 & \textbf{8,233.91} & 8,650.681 & \textbf{436.684}* & 1,098.085\\ \hline
        Validation MAE (\$) 					& \textbf{321.106} & 709.154 & 7,210.621 & \textbf{6,068.537} & \textbf{309.384}* & 708.485\\ \hline 
        Testing RMSE (\$) 						& 502.473 & \textbf{272.027} & 8,233.91 & \textbf{4,006.787} & 354.071 & \textbf{243.47}* \\ \hline
        Testing MAE (\$) 						& 321.106 & \textbf{207.332}  & 7,210.621 & \textbf{2,969.586} & 312.009 & \textbf{201.568}* \\ \hline
    \end{tabular}
    \caption{ \centering
Results of \emph{LSTM} and \emph{CoMForE} for different data modality combinations: (a) only trading data, (b) only sentiments, and (c) a combination of sentiments and trading data. The asterisk (*) highlights the lowest value for each modality, indicating the least error in each row.
}
    \label{tab:comp1}
\end{table*}

Additionally, we performed further experiments applying \emph{LSTM} and \emph{CoMForE} on trading data, supplemented with various other modalities; (d) the hash rate, (e) search volume, and (f) blockchain data encompassing all six modalities. The evaluation results of both models are presented in Table~\ref{tab:comp2}. While the error for (d) is marginally higher for \emph{CoMForE}, which is possibly due to the boosting process, the lower error observed for \emph{LSTM} when incorporating the hash rate suggests it as a beneficial factor for a more precise price forecasting. The findings from (e) suggest that search volumes did not notably influence the performance of either \emph{LSTM} or \emph{CoMForE} when combined with trading data. However, it can be speculated that search volumes could exhibit enhanced performance when integrated with other modalities, such as social media sentiments, as both represent a measure of ``public awareness''.

\begin{table*}[h!]
    \centering
    
    \begin{tabular}{|c||c|c||c|c||c|c|}
        \hline
        \multirow{2}{*}{} & \multicolumn{2}{c||}{(d) Trading + Hash rate} & \multicolumn{2}{c||}{(e) Trading + Search volume} & \multicolumn{2}{c|}{(f) Trading + Blockchain data}\\
        \cline{2-7}
         & \emph{LSTM} & \emph{CoMForE} 	& \emph{LSTM} & \emph{CoMForE} & \emph{LSTM} & \emph{CoMForE} \\\hline
        Training RMSE (\$) 						& 299.818 & \textbf{41.201}* & 352.823 & \textbf{89.093} & 280.944 & \textbf{96.525} \\ \hline 
        Training MAE (\$) 						& 178.795 & \textbf{14.433}* & 206.466 & \textbf{30.981} & 171.552 & \textbf{62.029} \\ \hline
        Validation RMSE (\$) 					& \textbf{433.68}* & 1,156.835 & \textbf{522.815}  & 1,438.16 & \textbf{1,052.821} & 1,151.279\\ \hline
        Validation MAE (\$) 					& \textbf{299.766}* & 782.773 & \textbf{369.701} & 940.658 & \textbf{707.377} & 765.778\\ \hline
        Testing RMSE (\$) 						& 433.680 & \textbf{356.554} & 519.175 & \textbf{277.861} & \textbf{200.263}* & 281.607\\ \hline
        Testing MAE (\$) 						& 299.766 & \textbf{291.09} & 365.0 & \textbf{201.177} & \textbf{156.694}* & 213.981\\ \hline
    \end{tabular}
     
    \caption{\centering  
    Results of \emph{LSTM} and \emph{CoMForE} for trading data combined with (d) the hash rate, (e) search volume and (f) blockchain data. The asterisk (*) highlights the lowest value for each modality, indicating the least error in each row.}
    \label{tab:comp2}
\end{table*}

\begin{figure}[h!]
    \centering
    \includegraphics[width=0.8\linewidth]{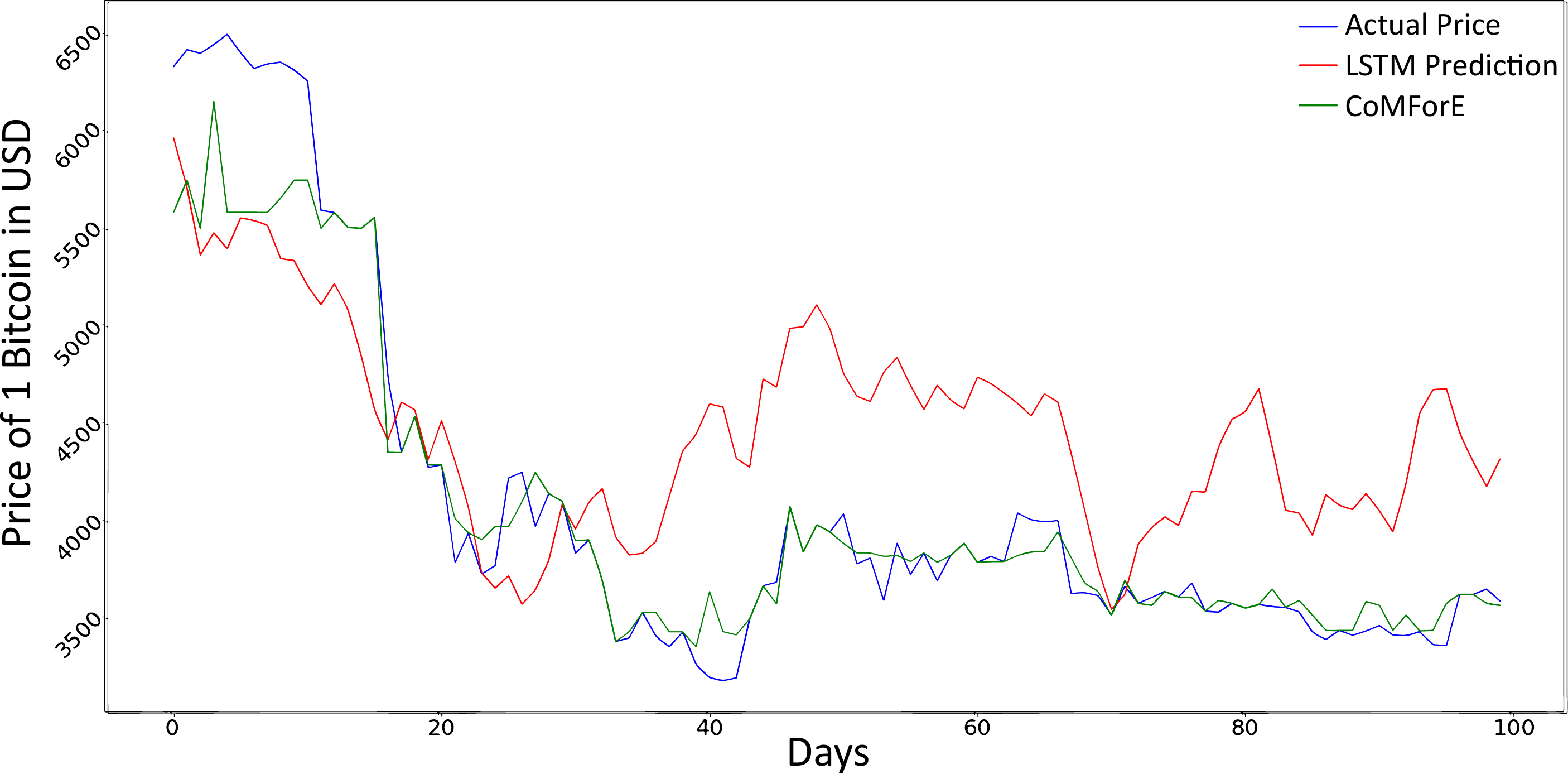}
    \caption{\centering \emph{LSTM} and \emph{CoMForE} forecasting results with blockchain data over 100-day interval.}
    \label{fig:onlyblockchain}
\end{figure}

\begin{figure}[h!]
\centering
    \includegraphics[width=0.85\linewidth]{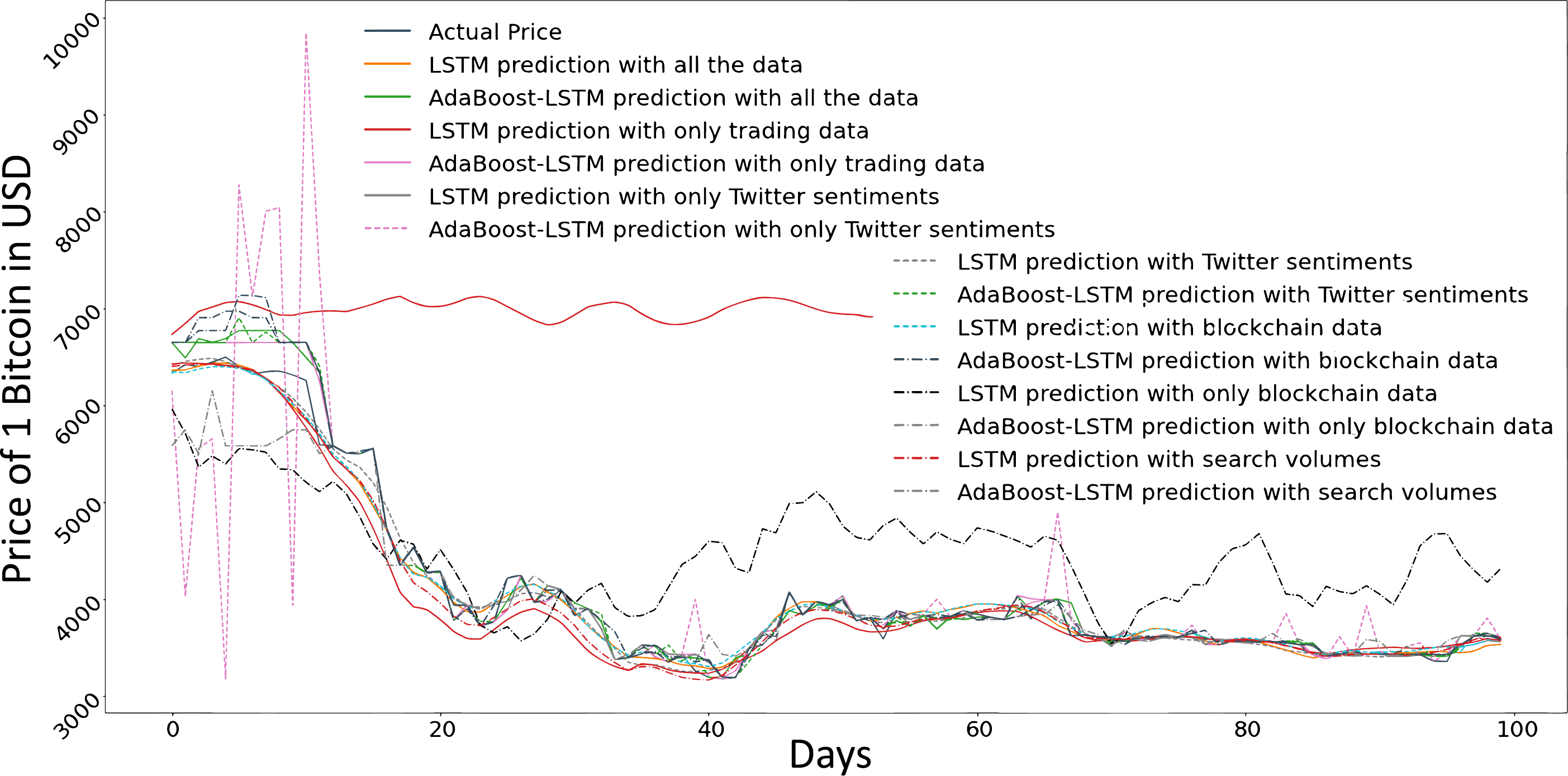}
    \caption{\centering \emph{CoMForE} and \emph{LSTM} outcomes across various modality combinations over 100-day interval.}
    \label{fig:allmodalitiesOneplot}
    \centering
\end{figure}

The outcomes presented in column (f) for both models substantiate that incorporating blockchain data and trading data significantly enhances predictive accuracy compared to using the hash rate alone. However, the testing errors encountered with the \emph{CoMForE} are notably larger than those found with the \emph{LSTM} model. As demonstrated in Figure \ref{fig:onlyblockchain}, when blockchain data was utilized as the sole input for both models, \emph{CoMForE} didn't precisely forecast the price, though it could track the price trend. These results highlight a noticeable correlation between blockchain data and cryptocurrency prices, which can assist in the prediction of cryptocurrency prices. To the best of our knowledge, this is the first study to integrate blockchain data into an analytical framework for predicting cryptocurrency prices. For further insights into the interplay between different modality combinations, Figure~\ref{fig:allmodalitiesOneplot} showcases the qualitative results of \emph{CoMForE} and \emph{LSTM} over a randomly sampled 100-day testing interval across all combinations.

\begin{figure}[h!]
\includegraphics[width=0.9\linewidth]{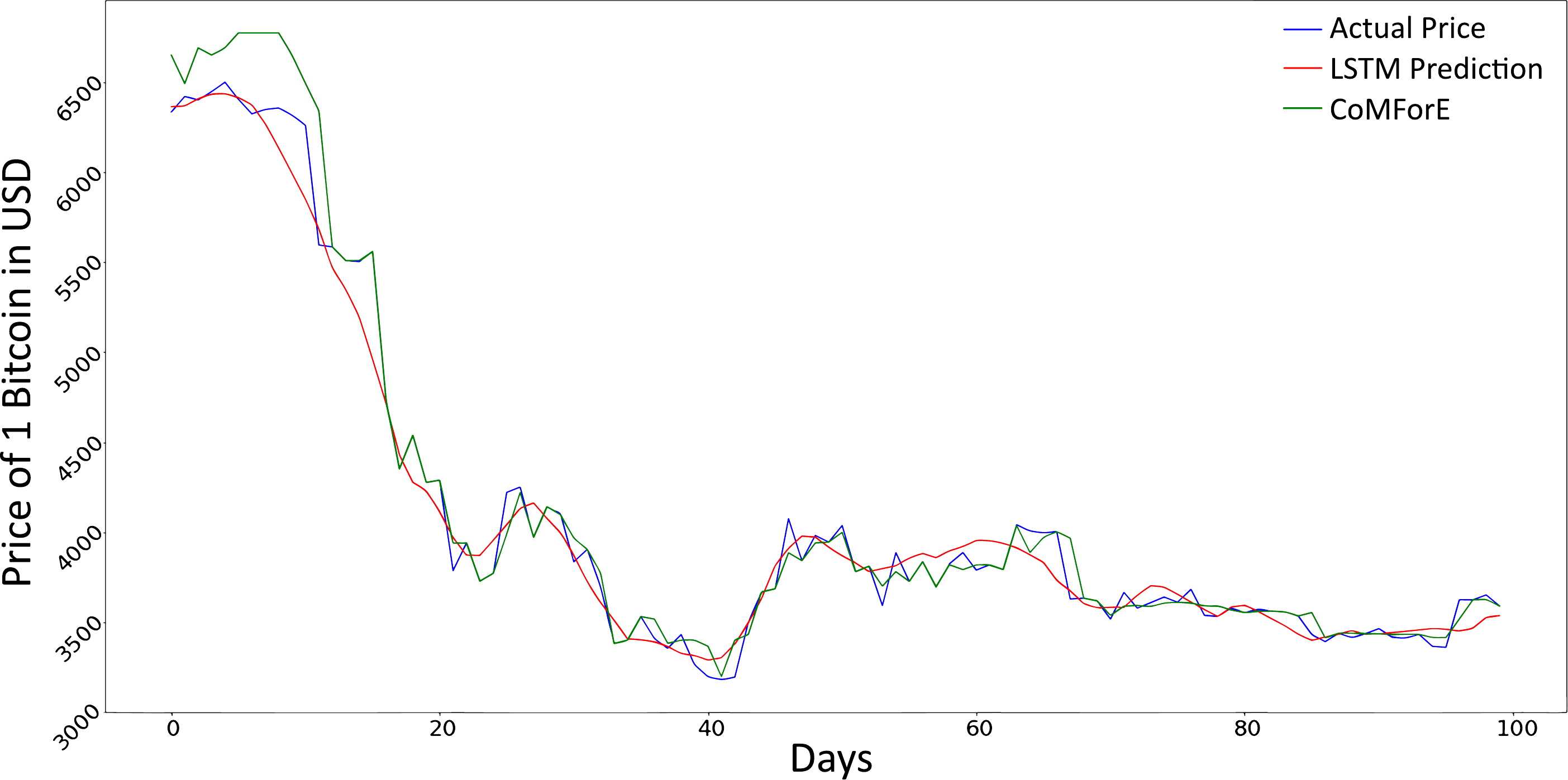}
\caption{ \emph{LSTM} and \emph{CoMForE} forecasting results with all modalities over 100-day interval.}
\label{fig:fullmodels}
\centering
\end{figure}

In our comprehensive evaluation, Table~\ref{tab:evaluationAll} presents the cumulative impact of combining all four modalities: trading data, social media sentiments, search volumes, and blockchain data. As clearly illustrated, this combination significantly enhances the results, indicating their complementary nature. Specifically, we observed an improvement of $\$75.312$ in MAE, which corresponds to a $36.32\%$ enhancement in forecasting accuracy. To visually corroborate this, Figure~\ref{fig:fullmodels} offers a qualitative comparison over a span of 100 days, where the forecasted trajectories from both models closely align with the actual price trends.

\begin{table}[h!]
\centering
    \begin{tabular}{|c|c|c|}
        \hline
         & {\centering \emph{LSTM} using All} & \multicolumn{1}{|p{2cm}|}{\centering \emph{CoMForE} using All}\\ \hline
        Training RMSE (\$) &  280.013 & \textbf{83.564} \\ \hline 
        Training MAE (\$) &  173.698 & \textbf{25.780} \\ \hline
        Validation RMSE (\$) &  389.245\$ & \textbf{234.718} \\ \hline
        Validation MAE (\$) &  281.399 & \textbf{234.666} \\ \hline
        Testing RMSE (\$) &  389.245 & \textbf{158.929} \\ \hline
        Testing MAE (\$) &  281.399 & \textbf{132.027} \\ \hline
    \end{tabular}

\caption{\centering Performance of \emph{LSTM} and \emph{CoMForE} with all modalities incorporated.}
\label{tab:evaluationAll}
\end{table}

In response to \textbf{RQ2}, Tables~\ref{tab:comp1}, \ref{tab:comp2}, and \ref{tab:evaluationAll} reveal the anticipated outcome, where \emph{CoMForE} shows substantial superiority over the \emph{LSTM} in all scenarios. In cases, (a) and (c), the \emph{LSTM}'s overall validation error appears lower, which could be due to the inconsistent trends and volatilities in price over time. Since \emph{LSTM} typically produces smoother forecasts than the weak learners of \emph{CoMForE}, it yields a lower error during periods of high volatility. As depicted in Figure \ref{fig:datasplit}, periods from 2012 to 2017 and the end of 2018 to 2019 exhibited minor fluctuations compared to other timeframes.

\begin{figure}[h!]
\centering
\begin{subfigure}[b]{\linewidth}
     \includegraphics[width=0.9\linewidth]{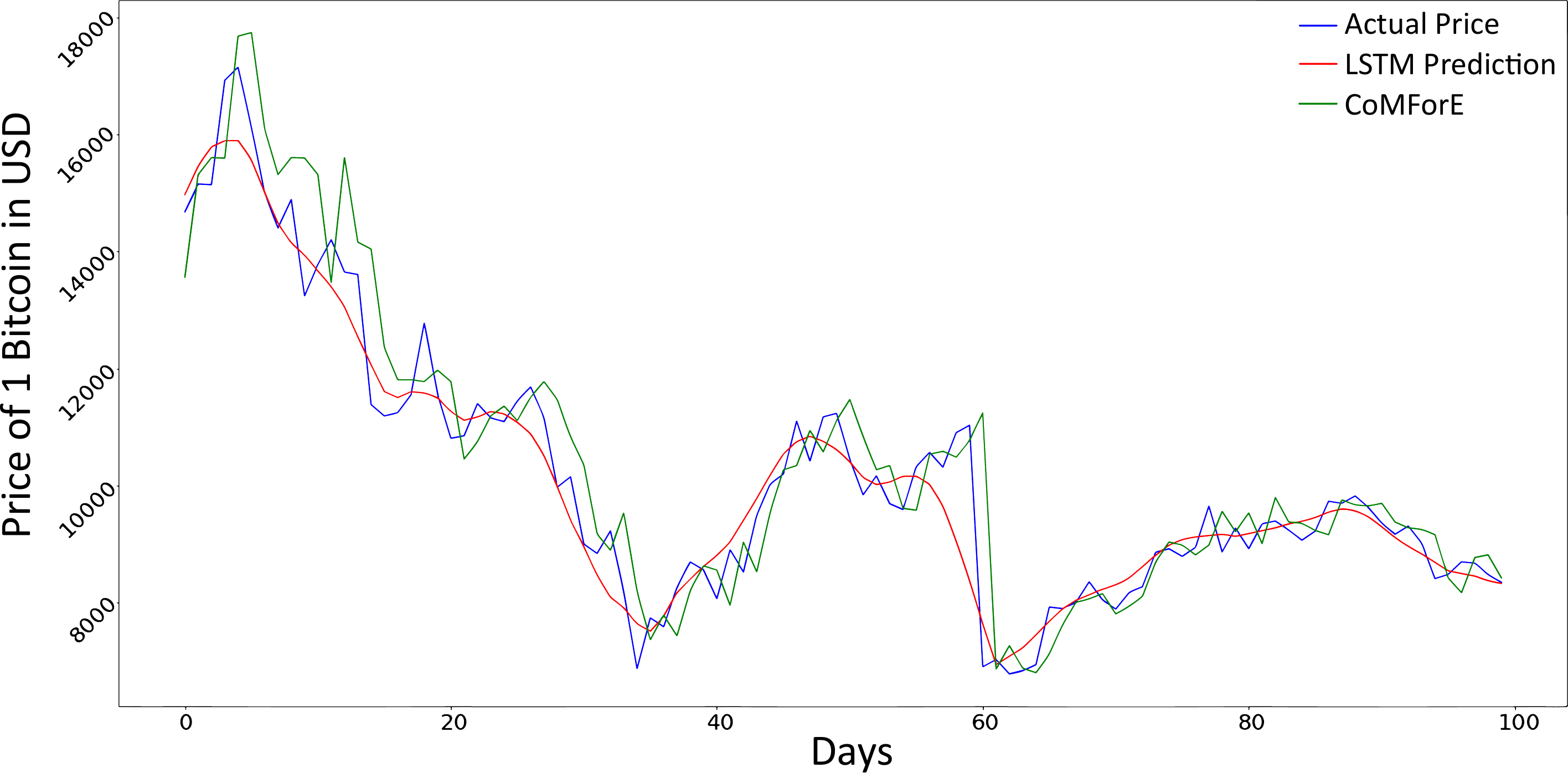}
     \caption{Only Trading}
     \label{fig:com1a}
\end{subfigure}
 \hfill
\begin{subfigure}[b]{\linewidth}
     \includegraphics[width=0.9\linewidth]{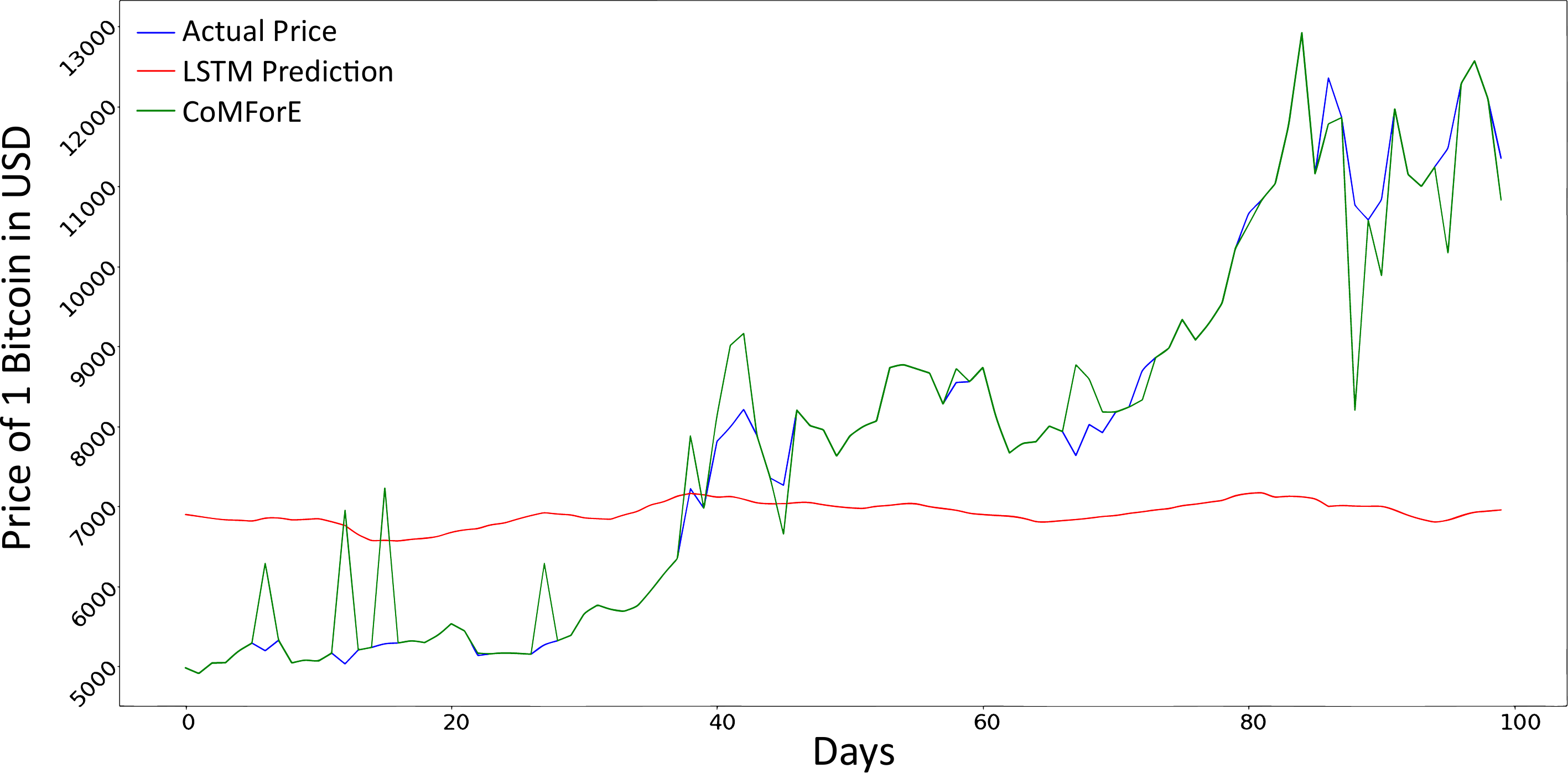}
     \caption{Only Sentiments}
     \label{fig:com1b}
\end{subfigure}

\begin{subfigure}[b]{\linewidth}
     \includegraphics[width=0.9\linewidth]{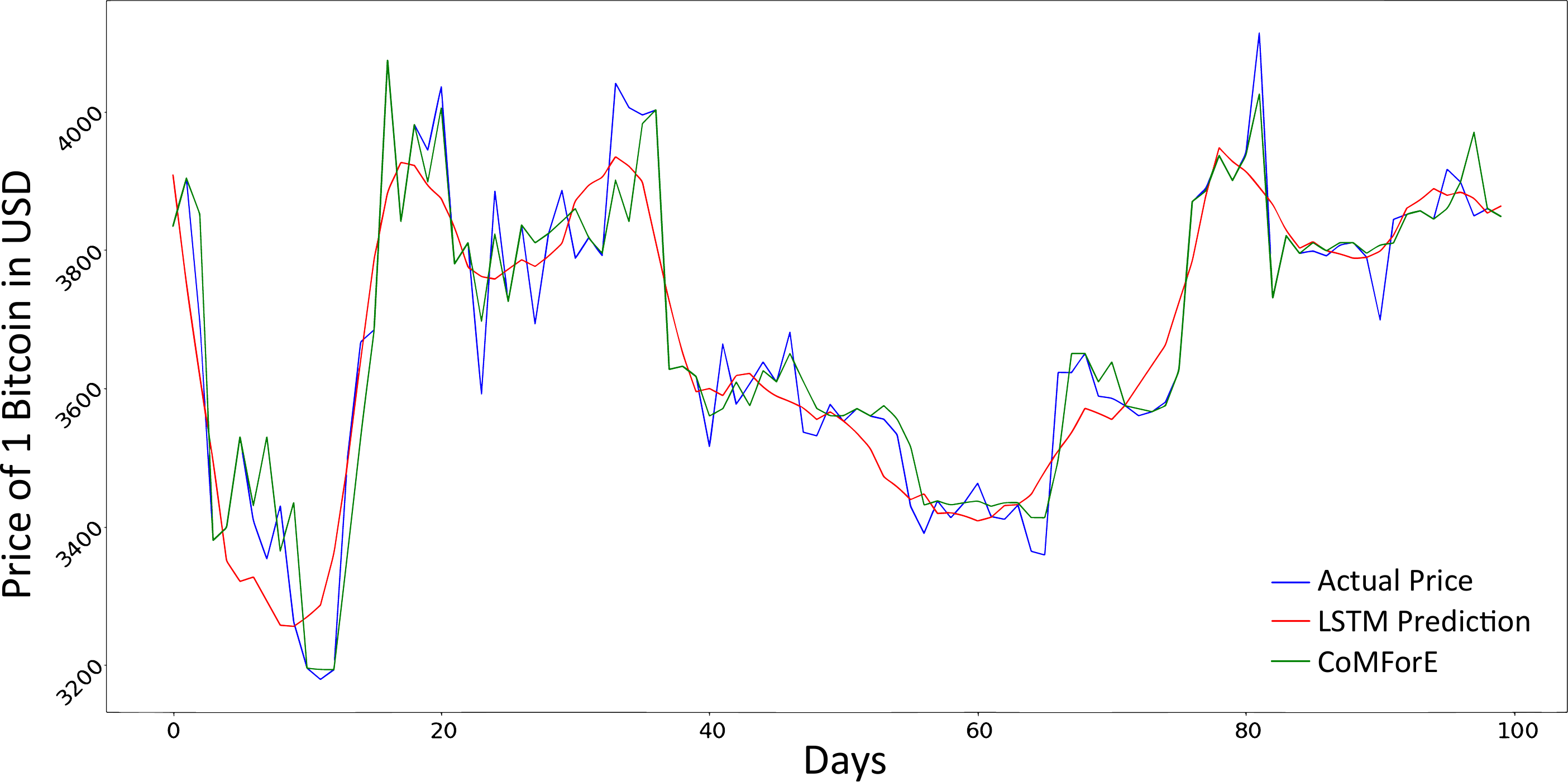}
     \caption{Trading + Sentiments}
     \label{fig:com1c}
\end{subfigure}

\caption{\centering Visual representation of the forecasting performance of \emph{LSTM} and \emph{CoMForE} over a 100-day interval, tested using (a) trading data, (b) sentiment data, and (c) a combination of both. }
\label{fig:com1}
    \label{fig:com1}
\end{figure}

\begin{figure}[h!]
\centering
\begin{subfigure}[b]{\linewidth}
\addtocounter{subfigure}{3}
\centering
     \includegraphics[width=0.8\linewidth]{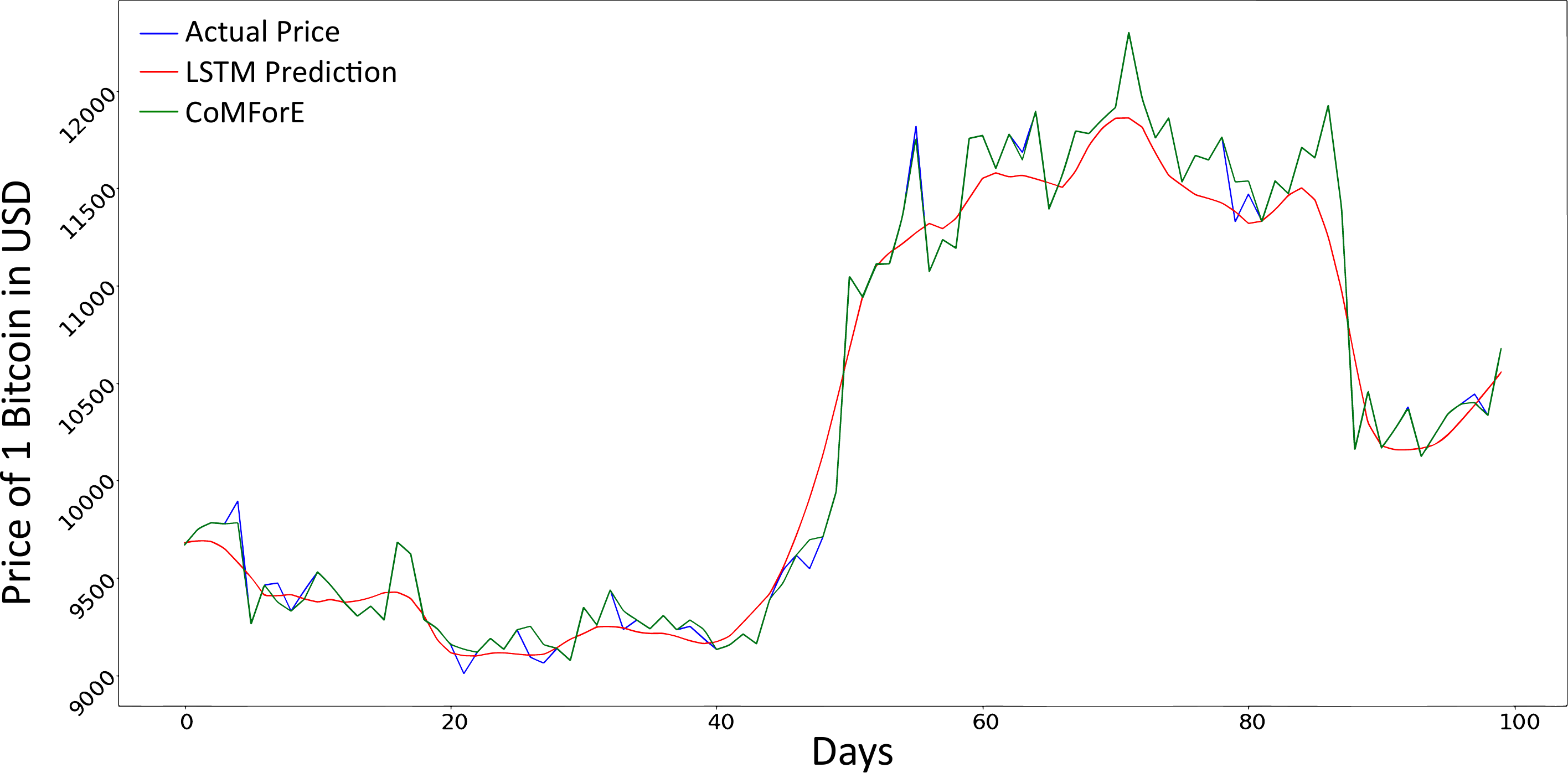}
     \caption{\centering Trading + Hashrate}
     \label{fig:com2a}
\end{subfigure}
 \hfill
\begin{subfigure}[b]{\linewidth}
\centering
     \includegraphics[width=0.8\linewidth]{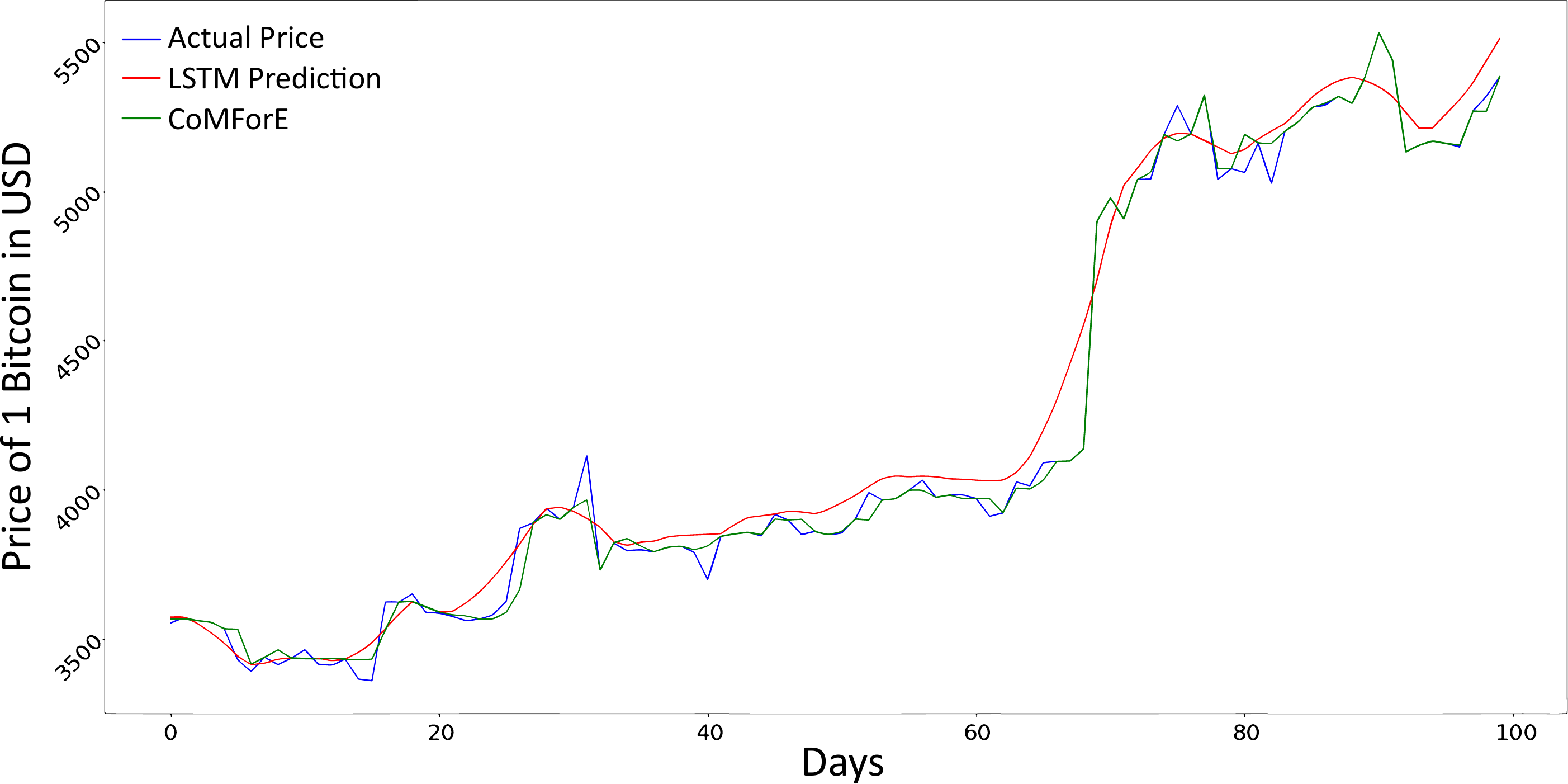}
     \caption{\centering Trading + Search volume}
     \label{fig:com2b}
\end{subfigure}

\begin{subfigure}[b]{\linewidth}
\centering
     \includegraphics[width=0.8\linewidth]{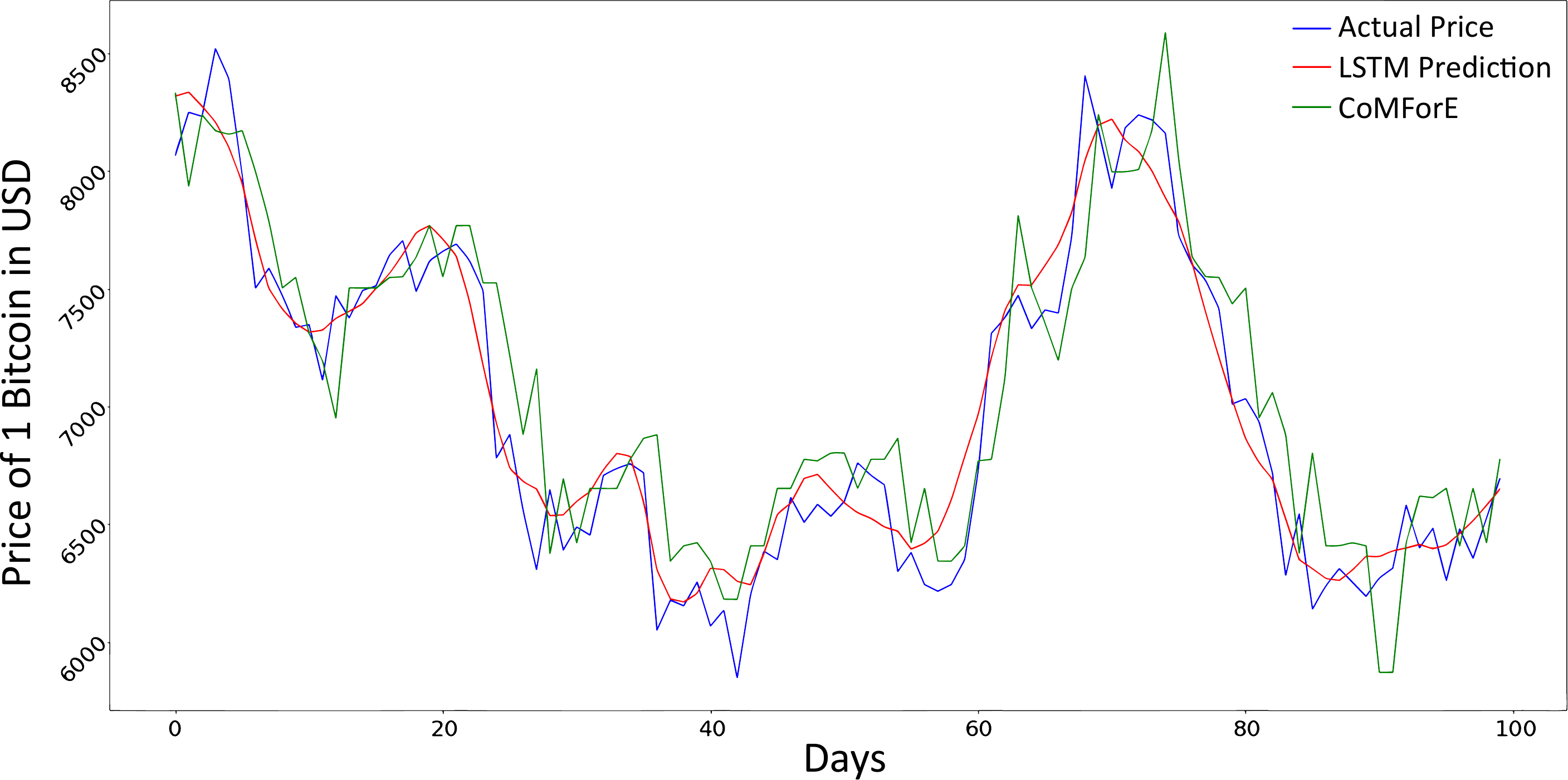}
     \caption{\centering Trading + Blockchain data}
     \label{fig:com2c}
\end{subfigure}

\caption{\centering  Visual representation of the forecasting performance of \emph{LSTM} and \emph{CoMForE} over a 100-day interval, tested using trading data combined with (d) the hash rate, (e) search volume, and (f) blockchain data.}
    \label{fig:com2}
\end{figure}

Figure \ref{fig:com1} presents the qualitative outcomes of the \emph{CoMForE} and \emph{LSTM} models over a randomly selected 100-day testing interval for combinations (a), (b), and (c). Similarly, Figure~\ref{fig:com2} illustrates the results for combinations (d), (e), and (f). In both figures, the blue curve represents the actual price. Meanwhile, the red and green curves depict the price forecasts generated by the \emph{LSTM} and \emph{CoMForE} models, respectively. Notably, \emph{CoMForE} consistently outperforms the single learner, \emph{LSTM}, especially for the combinations (a), (b), and (c). Although the forecasting error in (b) is notably higher than when only using trading data, it is significant to acknowledge that \emph{CoMForE}, despite being fed with a solitary input per day (i.e., Twitter sentiments), managed to detect the relationship between prices and sentiments, subsequently producing a meaningful price prediction. As illustrated in Figure \ref{fig:com1b}, \emph{CoMForE} successfully tracks the price trend using sentiment data alone. In contrast, the \emph{LSTM} model has difficulty learning patterns from the input data and tends to produce almost constant forecasts over time.

\begin{figure}[h!]
\centering
\includegraphics[width=0.9\linewidth]{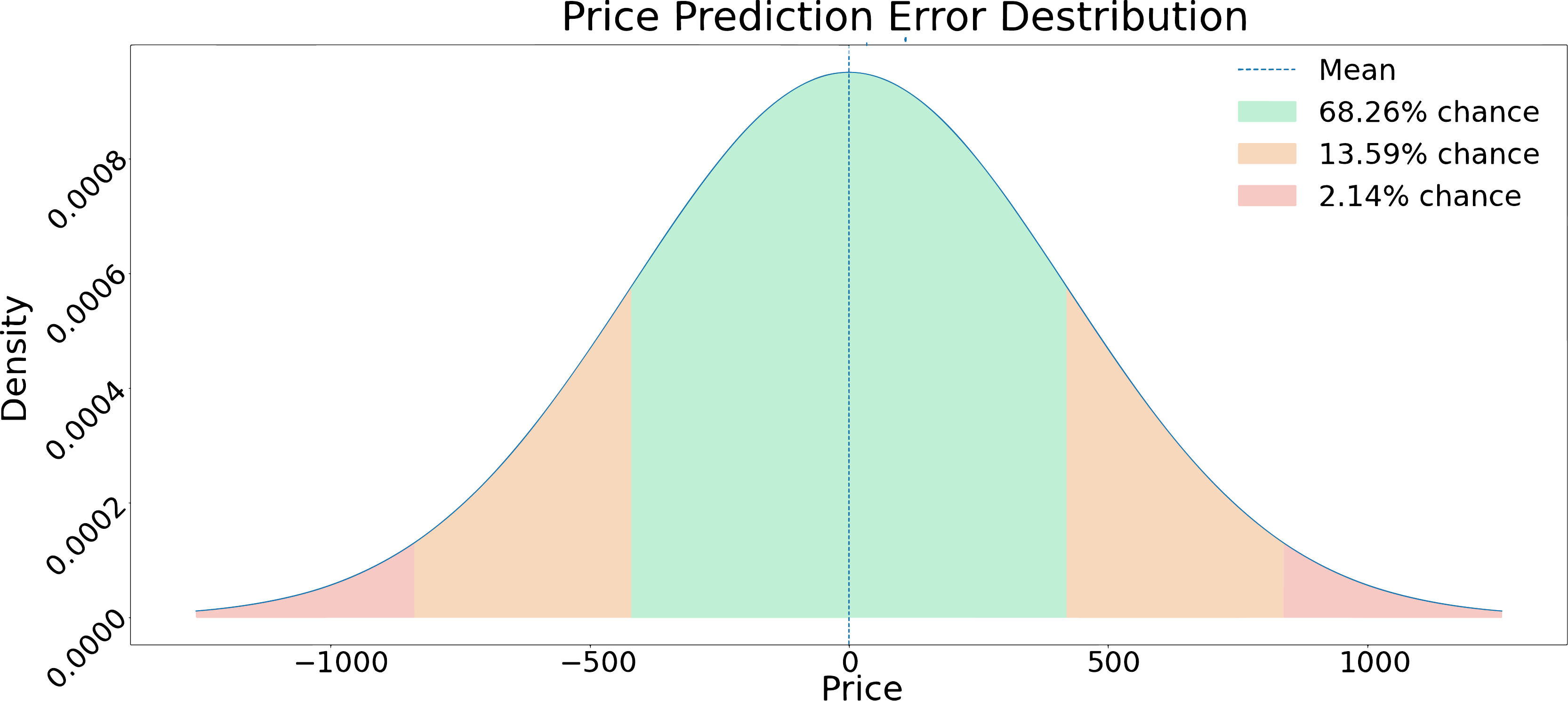}
\caption{MAE distribution over the entire testing set for \emph{CoMForE}.}
\label{fig:errordist}
\end{figure}

To gain a deeper understanding of the model's performance, we computed the \emph{MAE} distribution for the entire testing set as illustrated in Figure \ref{fig:errordist}. As can be seen, the forecasted price deviates by $+/- 500\$$ from the actual price of 1 \bitcoinA~ approximately 68\% of the time, yielding to an average error of $3.02\%$. This fluctuation is particularly significant given that the average \bitcoinA~ price within the testing set is \$16553.59. This deviation emphasizes the commendable precision of the model's forecasting relative to the actual price.

\subsubsection{Comparison against baseline approaches}
Additionally, we evaluate the performance of \emph{CoMForE}, trained with all modalities, in comparison to baseline methodologies. The comparison is conducted over the same dataset, within the period from January 1, 2020, to July 1, 2020. It is noteworthy that all models in this comparison are forecasting \bitcoinA~ prices for the subsequent 24 hours. As delineated by the results in Table \ref{tab:evaluationComparison}, it is evident that \emph{CoMForE} outperforms the other models, which capitalizes on multiple data modalities and on the power of ensemble learning, enhancing the understanding of the underlying trends and thereby improving the accuracy of its predictions.

\begin{table*}[ht!]
\centering
\begin{tabular}{|c|c|c|c|c|c|c|c|c|}
\hline
 & \emph{CoMForE} & \emph{BNN17} &  \emph{LSTM20A} & \emph{LSTM20B} & \emph{AdaBoost21} & \emph{GRU21} & \emph{ARIMA19} & \emph{GRU20} \\ \hline
RMSE (\$) & \textbf{158.929} & 221.642 & 166.875 & 202.787 & 197.544 & 208.006 & 584.384 & 185.154 \\ \hline
MAE (\$) & \textbf{132.027} & 184.124 & 148.628 & 177.02 & 156.346 & 160.44 & 542.73 & 157.632 \\ \hline
\end{tabular}
\caption{Evaluation Comparison Between the results of \emph{CoMForE} and baseline approaches on test data.}
\label{tab:evaluationComparison}
\end{table*}

\subsubsection{Fluctuation Analysis}
\label{apn:qlt}

\begin{figure}[h!]
  \begin{subfigure}[b]{\linewidth}
  \includegraphics[width=\textwidth]{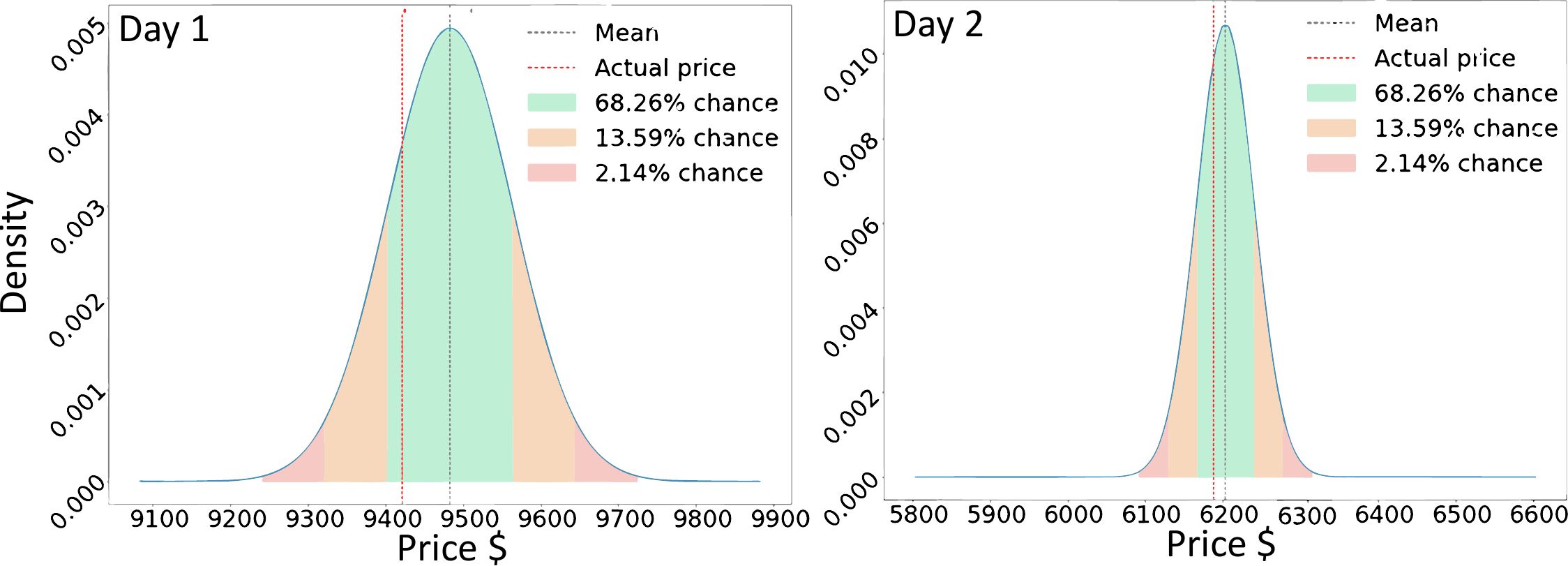}
  \end{subfigure}\\
  \begin{subfigure}[b]{\linewidth}
  \includegraphics[width=\textwidth]{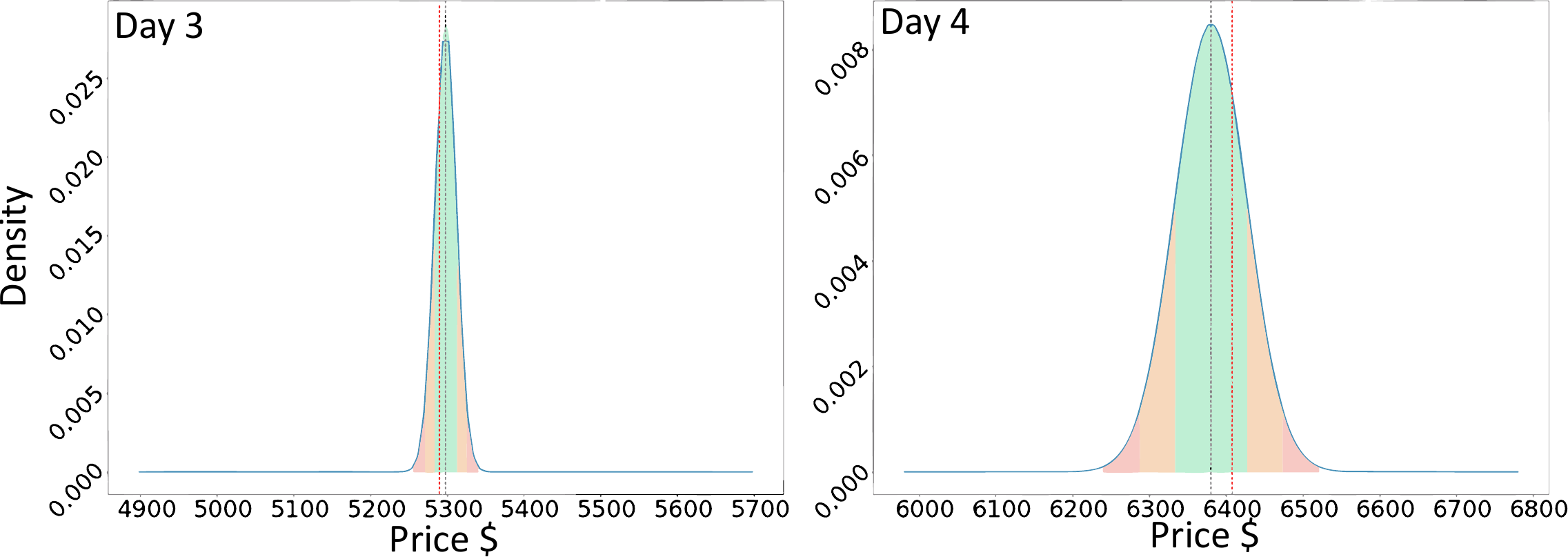}
  \end{subfigure}
    \caption{\centering Predicted price distribution of four randomly selected days. \emph{Mean} denotes the predicted price}
    \label{fig:priceDistributions}
    \centering
\end{figure}

In response to \textbf{RQ3}, we propose to equip the investor with not only the price forecast but also the range of potential price fluctuations. Given the complexity of quantifying this, we present in Figure \ref{fig:priceDistributions} a set of outcomes obtained using the approach described in Section~\ref{subsec:fa} over four random days. As illustrated, the certainty of price forecast can vary day-to-day. On days marked by significant volatility, the price distribution appears flat and exhibits a high standard deviation, indicating a high likelihood of deviation from the model's singular prediction. Conversely, on days when the model predicts with a high degree of certainty, the price distribution tends to peak sharply and display a lower standard deviation. These prediction distributions serve as a valuable resource for investors as they offer clear insights into the model's confidence in its predictions for a specific day.

\subsubsection{Long-term forecasting}
\begin{figure}[h!]
    \centering
    \includegraphics[width=0.85\linewidth]{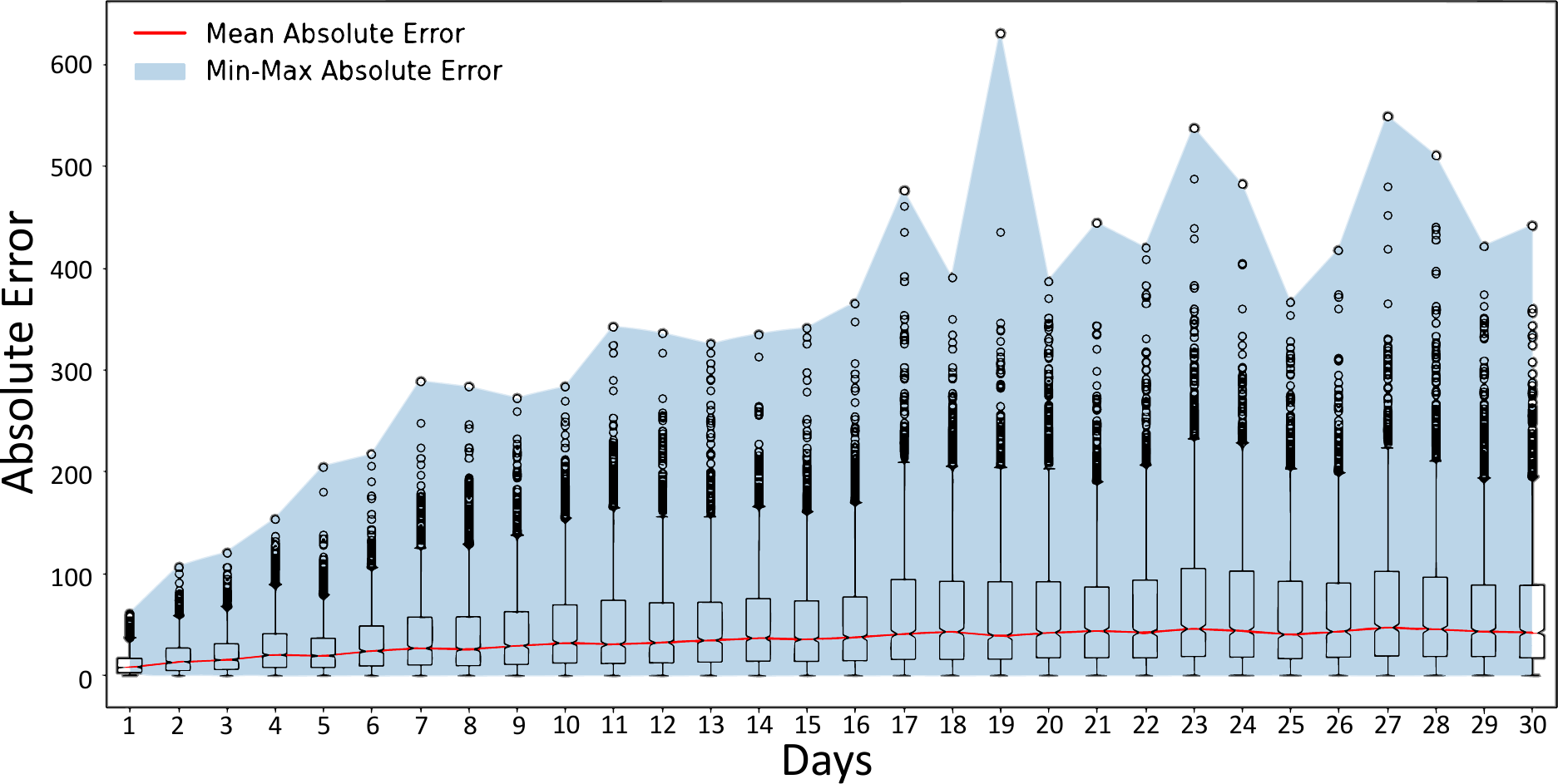}
    \begin{center}
            \caption{\centering Bitcoin price prediction absolute testing errors over different time windows}
            \label{fig:monthwindowtest}
    \end{center}
\end{figure}

In light of the promising obtained results, we further explored the limitations and performance of \emph{CoMForE} and multimodality combination in generating long-term predictions. Therefore, we continuously predicted the price by appending the last prediction to the input of the prediction of the subsequent day. This process is reiterated for up to the 30th day. It is crucial to note that only historical trading data is considered since other data modalities cannot be predicted. This assumption implies a likely performance dip because the model is model operating with less data for each successive day's prediction. Figure \ref{fig:monthwindowtest} showcases the results of this experiment, with the red curve representing the mean MAE of each time-window prediction. The boundaries of the blue area, meanwhile, indicate the highest and lowest MAE. As anticipated, the predictive accuracy is at its peak on the first day and progressively diminishes for longer-term forecasts.

\section{Conclusion}
\label{sec:con}

In this paper, we explored the impact of various independent factors on cryptocurrency volatility through a comprehensive multimodal forecasting approach, aiming to provide investors with accurate short-term price forecasts based on correlated sources (\textbf{RQ1}). The proposed approach harnesses the power of ensemble learning, combining multiple weak LSTM learners (\textbf{RQ2}. Understanding that investors need a more comprehensive view of the market, we introduced the concept of fluctuation distributions, which offer a broader perspective on the market and provide insights into the reliability of the price forecast (\textbf{RQ3}).

For future work, we plan to expand the application of this approach to other cryptocurrencies, incorporating the Bitcoin price as a vital variable. We also propose to generate and train the model on synthetic trading data characterized by extremely high volatility. This approach stems from our hypothesis that preparing the model for unprecedented market volatility will enhance its adaptability to future surprises inherent in the cryptocurrency markets.


\bibliographystyle{ACM-Reference-Format}
\balance
\bibliography{sample-base}

\appendix

\end{document}